# Interpreting core forms of urban morphology linked to urban functions with explainable graph neural network

Dongsheng Chen [a,*], Yu Feng [a], Xun Li [b], Mingya Qu [c], Peng Luo [d], Liqiu Meng [a]

[a] Chair of Cartography and Visual Analytics, Technical University of Munich, Munich 80333, Germany

[b] China Regional Coordinated Development and Rural Construction Institute, Sun Yat-sen University, Guangzhou 510275, China

[c] School of Tourism Management, Sun Yat-sen University, Guangzhou 510275, China

[d] Senseable City Lab, Massachusetts Institute of Technology, Massachusetts 02139, USA

**Abstract:** Understanding the high-order relationship between urban form and function is essential for modeling the underlying mechanisms of sustainable urban systems. Nevertheless, it is challenging to establish an accurate data representation for complex urban forms that are readily explicable in human terms. This study proposed the concept of **co**re urban **mo**rphology representation and developed an explainable deep learning framework for explicably symbolizing complex urban forms into the novel representation, which we call **CoMo**. By interpretating the well-trained deep learning model with a stable weighted F1-score of 89.14%, CoMo presents a promising approach for revealing links between urban function and urban form in terms of core urban morphology representation. Using Boston as a study area, we analyzed the core urban forms at the individual-building, block, and neighborhood level that are important to corresponding urban functions. The residential core forms follow a gradual morphological pattern along the urban spine, which is consistent with a center-urban-suburban transition. Furthermore, we prove that urban morphology directly affects land use efficiency, which has a significantly strong correlation with the location ($R^2$=0.721, $p<0.001$). Overall, CoMo can explicably symbolize urban forms, provide evidence for the classic urban location theory, and offer mechanistic insights for digital twins.

# 1. Introduction

The complex mechanisms of urban systems present a major scientific challenge in the construction of digital twins that address not only the physical entities but also their underlying mechanisms (Batty, 2024, 1971). Urban form and function represent two principal aspects of the operational dynamics of urban systems (Batty and Longley, 1994). Urban form refers to urban physical spaces (Balaian et al., 2024; Lynch, 1984) and includes three fundamental elements - buildings, blocks, streets, and their configurations (Fleischmann et al., 2022b). While urban function focuses on the land utilization for socioeconomic activities (Crooks et al., 2015). Urban form and function have a high-order relationship as urban form is both shaped by urban functions in a complex manner and determines how these functions evolve and how they are used by subjects (Besussi et al., 2010; Crooks et al., 2015). Based on the strong correlation, a series of studies have explored the prediction of urban functions by using urban morphological characteristics (Du et al., 2024; Xing and Meng, 2018; Yang et al., 2022). Understanding their high-order relationship has the potential to elucidate the mechanisms of urban systems, thereby supporting sustainable city systems (Kropf, 2018; Sun et al., 2024). However, the underlying relationship is still not clearly described and understood.

How do we quantitatively and accurately describe the high-order relationship between urban form and function? In the context of the growing development of city big data and artificial intelligence, an accurate data representation of complex urban forms has become a pivotal step toward the answer (Boeing, 2021; Janowicz et al., 2020). Urban form can be understood as a spatially symbolic system that affects and reveals operating modes of urban functional systems (Castex et al., 1980; Netto et al., 2023). Given a functional unit of an arbitrary shape, in order to quantitatively associate the corresponding relationship between function and form, the urban form with complex spatial information must be transformed into data representations. Urban form serves as the primary determinant of models of cities, which consist of their geometry and spatial configuration (Batty, 2024). Thus, the data representation of urban form relies on two components, i.e., spatial forms of the fundamental elements (individual level), and spatial patterns of the configurations of them within a given area (area level) (Fleischmann et al., 2022b). Regarding the first component, well-established studies of morphometrics have provided methodologies for quantifying the individual-level urban forms (Boeing, 2017; Fleischmann et al., 2022a, 2020), e.g., building morphology (Arribas-Bel and Fleischmann, 2022; Biljecki and Chow, 2022).

However, it remains a complex and challenging task to establish an accurate data representation for area-level urban form. Existing methods have explored three main types of data representation for urban form. 1) The statistic-based methods aggregate individual patterns into area-level ones via statistics (e.g., average, median, quartile) or metrics with limited spatial descriptive capabilities (Boeing, 2018; Fleischmann et al., 2022a). Thus, although the statistics are readily explicable in human terms, only limited information of area-level spatial patterns can be represented (Figure 1 (a)). 2) The advanced computer vision-based methods utilize grid-shape visual representation to portray figure-ground maps (Boeing, 2021; Wang et al., 2024; Wu and Biljecki, 2023). The novel visual representation focuses on the wholeness of areas and visual characteristics, thereby capturing rich information of area-level spatial patterns. However, they suffer from the fact that the surrounding irrelevant elements can creep in due to the inflexible grid-shape representation (Figure 1 (b)), which may impede the discernment of the relationship between morphology and function.

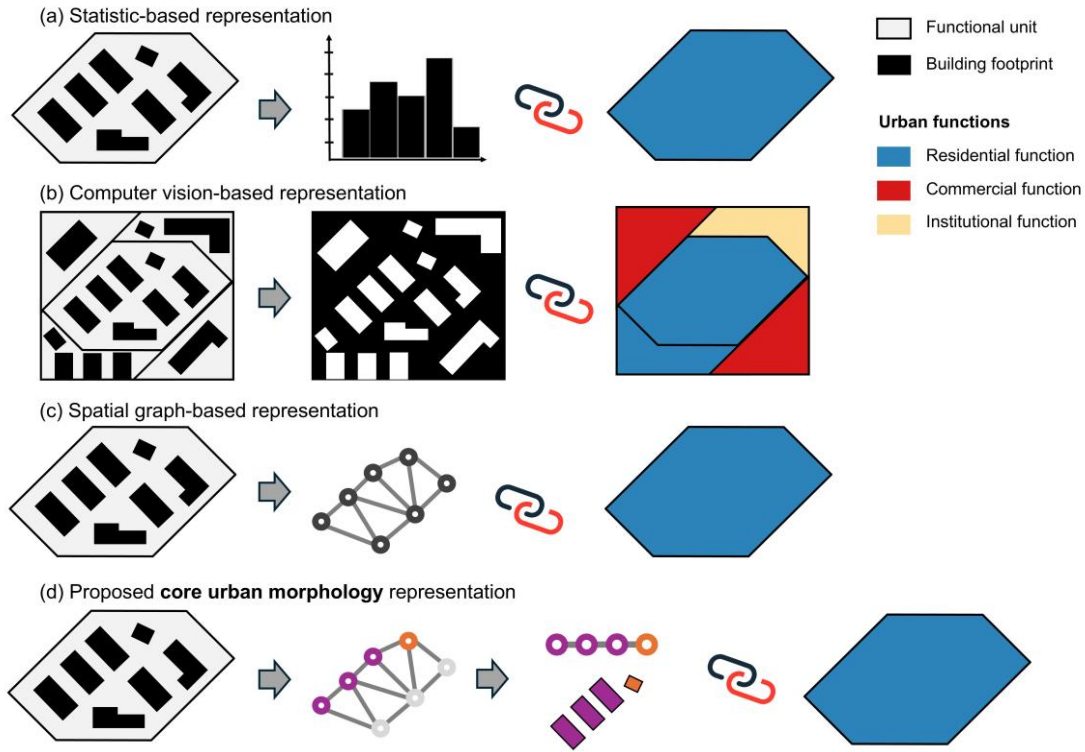

Figure 1 Concepts of the data representations for area-level urban forms, given a functional unit of an arbitrary shape. The proposed core urban morphology representation is capable of accurately characterizing complex urban forms and explicable in human terms.

3) The spatial graph-based representation can express the interconnectedness of urban elements with their neighbors (Figure 1 (c)), based on spatial geometry and proximity (de Almeida et al., 2013; Yan et al., 2019). Urban morphology focuses not only on urban elements, but also on the spatial interconnectedness of them (Wentz et al., 2018). Since the 1960s, a series of studies have developed various spatial graph-based representations (Szmytkie, 2017; Zagożdżon, 1970; Zhao et al., 2020). Recently, a popular integration of Delaunay triangulation, a classic spatial graph method (Jones et al., 1995), and graph neural networks (GNNs), an advanced graph-based deep learning methodology, has been proven powerful in accurately capturing characteristics of graphs (Yan et al., 2019; Zhao et al., 2020), with applications in urban function prediction (Kong et al., 2024; Yang et al., 2022). This effective pipeline provides a promising basis to accurately characterize area-level forms. Notwithstanding, interpretating spatial graphs and their links to urban functions is challenging due to the presence of rich spatial information of spatial graphs. This study proposes to convert the most representative subsets of spatial graphs into a spatially symbolic representation, which possess greater potential in describing the relationship to corresponding urban functions (Figure 1 (d)). Because these core parts with simplified, but the most important characteristics are more readily explicable in human terms. The spatially symbolic representation for urban morphology can be regarded as the core urban morphology representation linked to corresponding urban functions.

Explainable graph neural networks provide a potential means for identifying representative subsets of spatial graphs. Conventional GNNs usually treat a graph as an indivisible whole for the purpose of feature extraction due to its black-box nature, although excellent performance in prediction is shown (Li et al., 2023). Recently explainable GNN techniques have shown the potential to elucidate GNN model predictions for human understanding, thus supporting the analysis of the relationships between independent and dependent variables (Yuan et al., 2022). Amongst, GNNExplainer, as the most classic perturbation-based method, can identify important subsets of given graphs for corresponding predictions (Liu et al., 2023; Ying et al., 2019). It is a graph-masking approach by evaluating the importance of compact subgraphs and node features of graphs. It has the potential to identify representative subsets of urban forms, leading to the explicable analysis of the relationships to corresponding urban functions.

In this study, we propose a spatial graph-based explainable deep learning framework for symbolizing complex urban forms into the explicable **co**re urban **mo**rphology representation and analyzing the relationship to urban functions, called **CoMo**. Firstly, urban morphological graphs are formulated for each functional unit and input into a basic GNN model for linking the corresponding urban function. Secondly, the model-independent method GNNExplainer is proposed to be integrated to elucidate the most representative subsets of urban morphological graphs from the basic GNN model. Furthermore, an in-depth analysis is conducted to interpret complex urban forms in the form of core urban morphology representation. Boston, USA is taken as an illustrative case study to demonstrate the effectiveness of the representation. The discovered relationship allows for a more profound comprehension of the intrinsic logic of urban function systems, as well as the design of optimal urban forms.

## 2. Methodology

The proposed CoMo framework is based on urban morphological graph, graph neural network and graph explanation technologies (Figure 2). It includes three steps. 1) **Graph construction:** the urban morphological graphs are constructed for each functional unit based on morphometrics and Delauney Triangulation, in order to characterize the complex spatial connections between buildings and their spatial configurations. 2) **Learning:** a hierarchical graph pooling model combined with a graph convolutional network is taken as a basic GNN model of predicting urban functions via urban morphological graphs. A well-trained GNN model is regarded as the AI mastering the relationship between urban morphology and function. 3) **Interpretation:** CoMo interprets the basic GNN model to identify representative subsets of urban morphological graphs associated with corresponding urban functions via GNNExplainer, and then symbolizes them into core urban morphology representation. Finally, the high-order links between urban forms and urban functions are analyzed and explained via core urban form patterns.

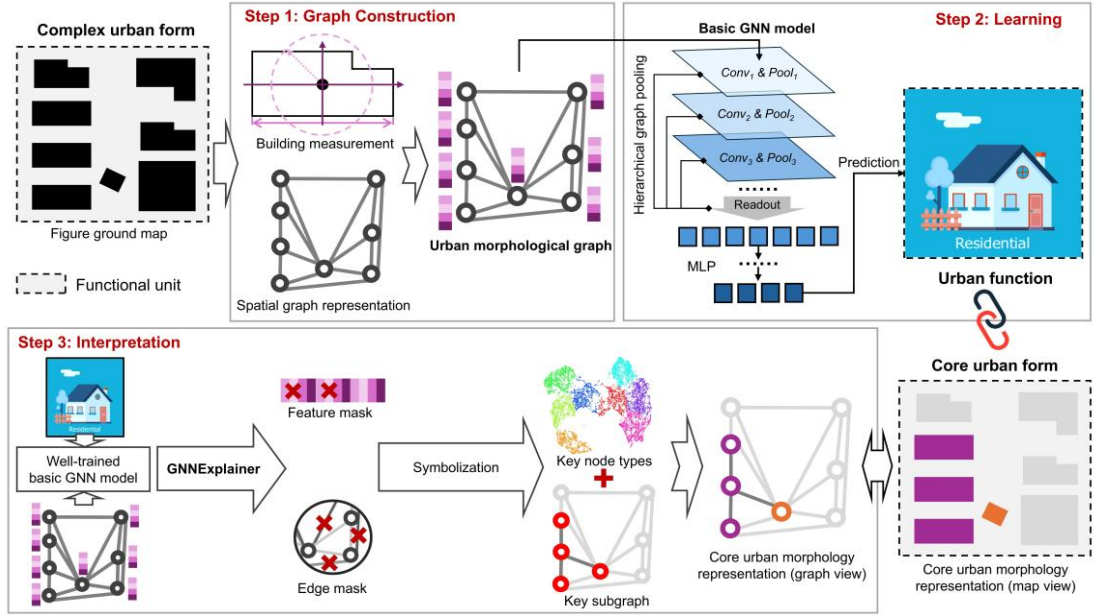

Figure 2 Workflow of the proposed CoMo framework.

## 2.1 Construction of urban morphological graphs

### 2.1.1 Morphological measurement for individual building-level forms

This step aims to quantitatively characterize diverse urban forms based on urban morphometrics in the building level. Its success can facilitate subsequent processing. Table 1 presents the equations and descriptions of the representative morphometrics that we have carefully selected from a vast array of morphological indicators based on their relative ease of interpretation. They cover four dimensions of building morphology, i.e., size, shape, orientation, and density. Four indicators are used to express the building size, i.e., area, perimeter, length of the longest chord, and mean radius. Four other indicators are used to describe the orientation of building polygons, particularly those complex ones. Shape is the most important dimension that characterizes the morphology of individual buildings. Thus, we collected 13 indicators to portray the diverse polygon shapes, including complexity, compactness, fractality, etc. To represent the density of the buildings, the area ratio of the building to its tessellation is applied. Tessellation cell, a kind of spatial unit in urban community planning, refers to a building's impact areas, including its footprint and the surrounding open space (Fleischmann et al., 2020). The density metric reveals the built intensity and proportion of urban open space, which is an important part of urban morphology.

Table 1 Morphometrics for building-scale urban form

| Dimension | Metric | Equation | Description |
|---|---|---|---|
| Size | Area (A) | - | Area of a polygon |
| | Perimeter (P) | - | Length of an outer perimeter wall of a polygon |
| | Longest chord's length ($L_{max}$) | - | Length of the longest chord in the convex hull of a polygon |
| | Mean radius ($\bar{r}$) | $\frac{1}{n}\sum_{i=1}^{n} r_i$ | Average distance from the polygon vertexes to its centroid |
| Orientation | SBRO | - | Orientation of the smallest bounding rectangle (SBR) of a polygon (Yan et al., 2019) |
| | Longest chord's orientation (LCO) | - | Orientation of the longest chord in the convex hull of a polygon |
| | Bisector orientation | - | Orientation of the bisector in the convex hull of a polygon |
| | Weighted orientation | $\frac{\sum_{i=1}^{n} edge_i \times Ori_i}{\sum_{i=1}^{n} edge_i}$ | Orientation representing the dominant direction where the longest edges of a polygon have the most significant alignment (Duchêne et al., 2003) |
| Shape | Complexity index | $\frac{A}{P}$ | Ratio of the area and the perimeter |
| | Iso-perimetric quotient (IPQ) circularity | $\frac{4\pi A}{P^2}$ | Measure of quadratic relationship between the area and the perimeter (Li et al., 2013a) |
| | Fractality | $1 - \frac{\log A}{2\log P}$ | Logarithmic relationship between the area and the perimeter (Basaraner and Cetinkaya, 2017) |
| | Max circularity | $\frac{\sqrt{A/\pi}}{r_{max}}$ | Ratio of the radius of the equal area circle to the longest radius of a polygon (Smith, 2014) |
| | Gibbs compactness | $\frac{4A}{\pi L_{max}^2}$ | Area ratio of a polygon to an ideal shape (Gibbs, 1961) |
| | Elongation index | $\frac{L_{SBR}}{W_{SBR}}$ | Length-width ratio of the SBR of a polygon |
| | Ellipticity | $\frac{L_{width}}{L_{max}}$ | Length ratio of the longest chord's perpendicular line to the longest chord |
| | Concavity | $\frac{A}{A_{CH}}$ | Area ratio of a polygon to its convex hull |
| | Digital compactness measure (DCM) | $\frac{A}{A_{SCC}}$ | Area ratio of a polygon to the smallest circumscribing circle (SCC) (Li et al., 2013b) |
| | Exchange index | $1 - \frac{A \cap A_{EAC}}{A_{EAC}}$ | Area ratio of the intersection between the building and its equal area circle (EAC) to the circle |
| | Boyce-Clark shape Index (BCI) | $\sum_{i=1}^{n} \lvert \frac{r_i}{\sum_{i=1}^{n} r_i} - \frac{100}{n} \rvert$ | Radii are drawn from a central point to the polygon's edges, compared to a standard circle (Zhang et al., 2023) |
| | Moments of inertia ($\mu 11$) | $\sum_{i=1}^{n} x_i \times y_i$ | Variance (spreading) around centroid in x and y of building vertexes (Jähne, 2013) |
| | Eccentricity | $\frac{r_{max}}{r_{min}}$ | Ratio of longest to shortest distance vectors from the polygon's centroid to its boundaries (Awcock and Thomas, 1995) |
| Density | Covered area ratio | $\frac{A}{A_{Tes}}$ | Proportion of a tessellation covered by a polygon (Fleischmann et al., 2020) |

### 2.1.2 Spatial graph representation for functional unit-level forms

Given any functional unit, an urban morphological graph is constructed in the representation of $g = (V, E, F, W)$. The centroids of the buildings within the functional unit are taken as $V$, the vertices of the urban morphological graphs, while the morphological features are regarded as attributes of the vertices $F$. In order to illustrate the interconnections between a given building and its four neighboring edifices, constrained Delauney triangulation is constructed (Yan et al., 2019; Zhao et al., 2020), showing undirected edges of the urban morphology graphs, which is represented as an adjacent matrix $E$. Delauney triangulation can link vertices with their neighbor vertices and has found applications in urban morphology and urban region representation. And the geographic distances between the buildings are regarded as the weights of the edges $W$. For $n$ functional units of an entire city, a set of urban morphological graphs can be constructed as $G = \{g_1, g_2, ..., g_n\}$.

Since this study applies block-level urban functional data, block is taken as the function unit to ensure the well-defined functional category of each unit is aligned with its respective urban form (Yang et al., 2022). It is worth noting that other levels of units can also be set as functional units as long as the functional categories of urban functional data are well defined. Blocks comprising fewer than four buildings are considered to lack a regional-scale spatial pattern, exhibiting instead a configuration of individual building patterns. Furthermore, the construction of a Delauney triangulation necessitates the presence of at least four nodes. Accordingly, this study focuses exclusively on blocks containing a minimum of four buildings.

## 2.2 Basic GNN model for urban function prediction

The step aims to generate a basic GNN model for deep learning of the relationship between urban functions and the constructed urban morphological graphs, $U = f(G, F)$, where $U$ represents the types of urban functions and $f$ represents the basic GNN model for predicting urban functions. This study adopted a hierarchical graph convolutional network (GCN) as the urban function prediction model, which consists of two parts - hierarchical graph pooling and a readout function (Figure 3).

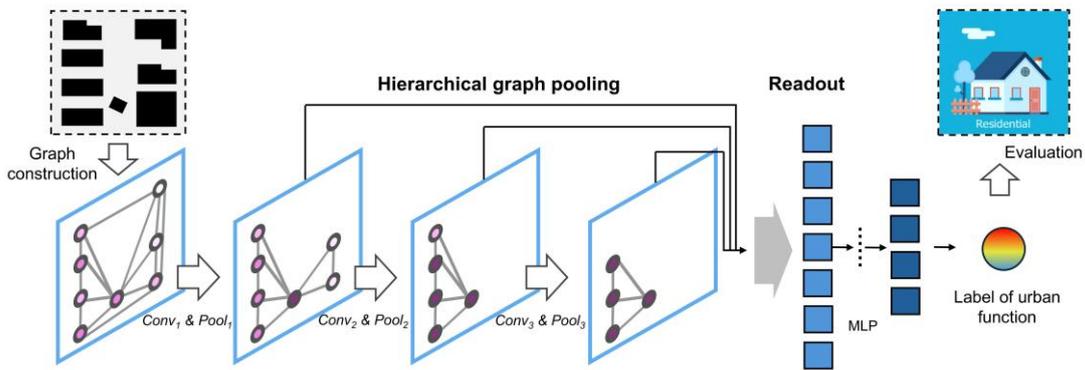

Figure 3 Architecture of the GCN model for urban function prediction via urban morphology

The part of hierarchical graph pooling contains multiple layers with graph convolutional operations and graph pooling operations. Convolutional operations aim to transfer vertex features and edge weight information of surrounding vertices to the target vertex. While pooling operations select the most informative vertices to form a smaller graph using an information criterion called vertex information function $\gamma$:

$$c_i = \gamma(g_i) = \left\| (I_i^j - (D_i^j)^{-1} E_i^j) F_i^j \right\|_1 \tag{1}$$

where $c, I, D$ refers to the list of vertex information scores, the identity matrix and the diagonal degree matrix of $E$. $\|\cdot\|_1$ is a row-wise L1 norm function. And $i, j$ refers to the $i$-th functional unit and the $j$-th layer in the hierarchical graph pooling part. And then the smallest graph in the last layer is converted into graph-level representations via the concat pooling operation combining mean-pooling and max-pooling.

The part of the readout function is for aggregation of graph-level representations to classify the input graph. A Multi-Layer Perceptron (MLP) with softmax layer is applied as the readout function, which can stably convert the graph-level representations into the embeddings with a fixed size. And the softmax layer performs the graph classification in the final step.

Based on the previous experiments (Zhang et al., 2019), the hierarchical GCN has 3 convolutional layers, 3 hierarchical pooling layers and 3 linear layers. The pool rate is set to 30% and the dimension of hidden representations is 64. Regarding the training parameters, batch size is set to 64 while the learning rate is set to 0.001. The dataset is a split into a training set, a validation set and a test set in a ratio of 0.8:0.1:0.1. And the basic GNN model is trained in a maximum of 1000 epochs. An early stop strategy was implemented to circumvent model overfitting, whereby the training process would be terminated if the accuracy on the validation set did not show improvement over a consecutive 70 epochs.

**Accuracy assessment:** In view of the naturally existing imbalanced class distribution in this study, a range of 10 different metrics were utilized to comprehensively evaluate the model performance, including Micro Precision, Micro Recall, Mirco F1-score, Macro Precision, Macro Recall (Balanced accuracy), Marco F1-score, Weighted Precision, Weighted Recall, Weighted F1-score, and Overall accuracy (OA) (Grandini et al., 2020). The equations and descriptions for the metrics are shown in Table S1 (see *Supplementary Materials*). Furthermore, the k-fold cross validation procedure with stratified random sampling was incorporated into the accuracy assessment (Prusty et al., 2022). K-fold cross validation evaluates the generalization and the stability of the model to ensure that the model has a consistent performance in each fold. While stratified k-fold ensures that the proportion of each class is approximately the same in all folds. Since the ratio of the test set was set to 0.1, 10 folds were established to cross-validate the whole dataset.

**2.3 Interpretation via core urban morphology representation**

This step aims to introduce the explainable graph neural network and an analysis framework to generate the spatially symbolic representation for core urban forms, i.e., the core urban morphology representation, for interpreting their relationships to corresponding urban functions.

2.3.1    Explainable graph neural network

Given a well-trained $f(G, F)$ model for predicting urban functions $U$, GNNExplainer is utilized to identify the connected subgraphs $G_{Si}$ with the relevant morphological features $F_{Si}$ that are key for predicting the urban function $u_i$. GNNExplainer is a model-agnostic approach that can explain the prediction of a black-box graph neural network based on mutual information *MI* formulated as follows (Ying et al., 2019).

$$max_{G_S} MI(U, (G_S, F_S)) = H(U) - H(U|G = G_S, F = F_S) \quad (2)$$

where *MI* indicates the change in the probability of $U = f(G, F)$ when the input is limited to subgraph $G_S$ and a subset of the vertex features $F_S$. $H$ refers to an entropy term. $G_S$ are connected subgraphs wherein the building vertices cover at most a one-hop region. $G_S, F_S$ are selected based on a comparison of the average importance generated by GNNExplainer and the designed global importance thresholds. Empirically, the feature threshold is set to 0.950 while the edge threshold is set to 0.900. From the perspective of urban morphology, features $F_S$ represent the key morphometrics of buildings while subgraphs $G_S$ refer to the core configurations of buildings.

### 2.3.2 Symbolization and interpretation of core urban forms

This study proposes a bottom-top analysis framework to convert the subgraphs and the features generated by GNNExplainer into the explicable core urban morphology representation. It includes three progressive steps involving three spatial levels, i.e., individual-building level, functional-unit level and regional level (Figure 4). In our case, the functional-unit level is set at the block level to keep consistent with the basic spatial unit of urban functions. And the regional level is set at the neighborhood level for better analysis and visualization of the city-wide spatial structure.

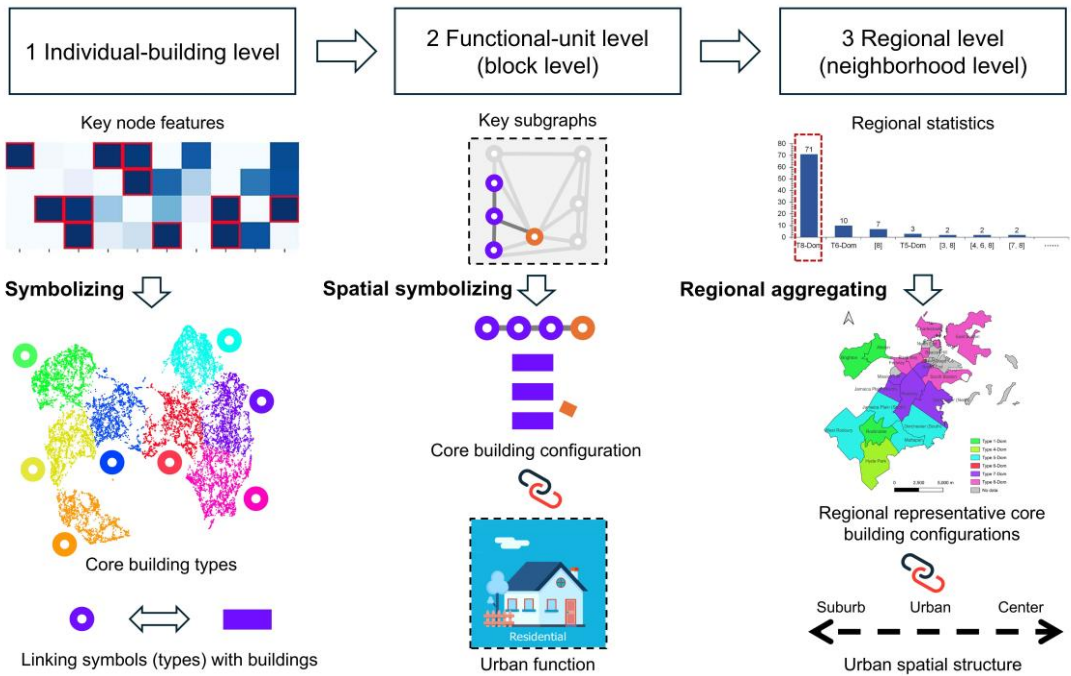

Figure 4 Framework of symbolizing and interpreting core urban forms.

**Step 1:** In the individual-building level, our framework proposes to conduct dimension reduction and clustering techniques to symbolize the key node features $F_S$. Firstly, the uniform manifold approximation and projection (UMAP) (McInnes et al., 2020) as well as the K-Means (Krishna and Murty, 1999) method are applied to symbolize the morphology of buildings. UMAP, a state-of-the-art tool for dimension reduction, excels at transforming high-dimensional features into low-dimensional representations while preserving the essential structure and relationships (Becht et al., 2019), thereby enhancing the interpretability. In this study, UMAP convert $F_S$ into two-dimensional representations. And then a clustering algorithm is used to further categorize the morphological types of buildings. The selection and settings of clustering algorithms are typically based on the characteristics of the data itself and the evaluation of clustering quality (Rubiños et al., 2024; Xu and Wunsch, 2005). Comparative experiments were performed (see Section S2 in *Supplementary Materials*) and K-Means (K=8) is confirmed as an appropriate clustering method for the data in this study. Finally, the morphological types are linked to the corresponding core buildings to assign morphological meanings. It' worth noting that CoMo is flexible enough to use other dimension reduction and clustering techniques based on characteristics of the data.

**Step 2:** In the functional-unit level, the morphological types of core buildings are firstly assigned to the key subgraphs $G_S = \{T_x^1, T_x^2, \ldots, T_x^n\}$. $T_x^n$ refers to the morphological Type $x$ for the *n*-th buildings. Based on the majority of morphological types, building configurations $G_S$ is converted into a symbolic representation, including two main categories of core building configurations, i.e., dominant type and diverse type. The former refers to the building configurations that consist of a dominating morphological type of core buildings beyond a given dominance threshold while the latter does not exhibit a main morphological type. In particular, the dominant type of configuration can be further categorized as a T$x$-dom configuration if the Type-$x$ is a dominating type. In this study, the general dominance threshold is set as 50% and the dominant type has not less than 2 buildings. While the threshold for subgraphs in residential function is set more strictly as 60% since residential areas usually have more buildings than the other areas. The implementation of a stricter threshold may facilitate a clearer presentation of the statistics associated with the diverse residential configurations. Finally, the types of core building configurations, representing the block-level core urban forms, are linked to their corresponding urban functions.

**Step 3:** In the regional level, a spatial statistical analysis was conducted for the frequency of different types of core building configurations within a region. The most frequent type of core building configurations in a neighborhood is defined as the regional representative type of configurations of the neighborhood. The regionally aggregated results can reveal a geographic distribution in the whole city. Through this geographical distribution, the relationship between the core urban forms of the neighborhood level and the urban spatial structure is analyzed. In this study, this step is specific for the residential function since sufficient residential samples are distributed in city space.

## 3. Results

### 3.1 Implementation

#### 3.1.1 Study area and data collection

Boston, USA, is taken as our study area (Figure 5). With its origins dating back to 1630, Boston has evolved from a colonial settlement to the capital city of Massachusetts and the most populous city in the state. The form of Boston started from the organic style with narrow, winding streets and irregular plots. Since 1882, Boston had grown into an industrialized city in terms of the urban design paradigm, which led to a more planned and regular grid street pattern. The urban spatial structure of modern Boston shows a concentric zone pattern centered on the downtown and surrounding districts. Boston offers a rich historical context of urban morphology and serves as one of the case study cities in Kevin Lynch's seminal work, "The Image of the City" (Lynch, 1964). Thus, it is a representative study area for us to explore the relationship between urban form and urban function.

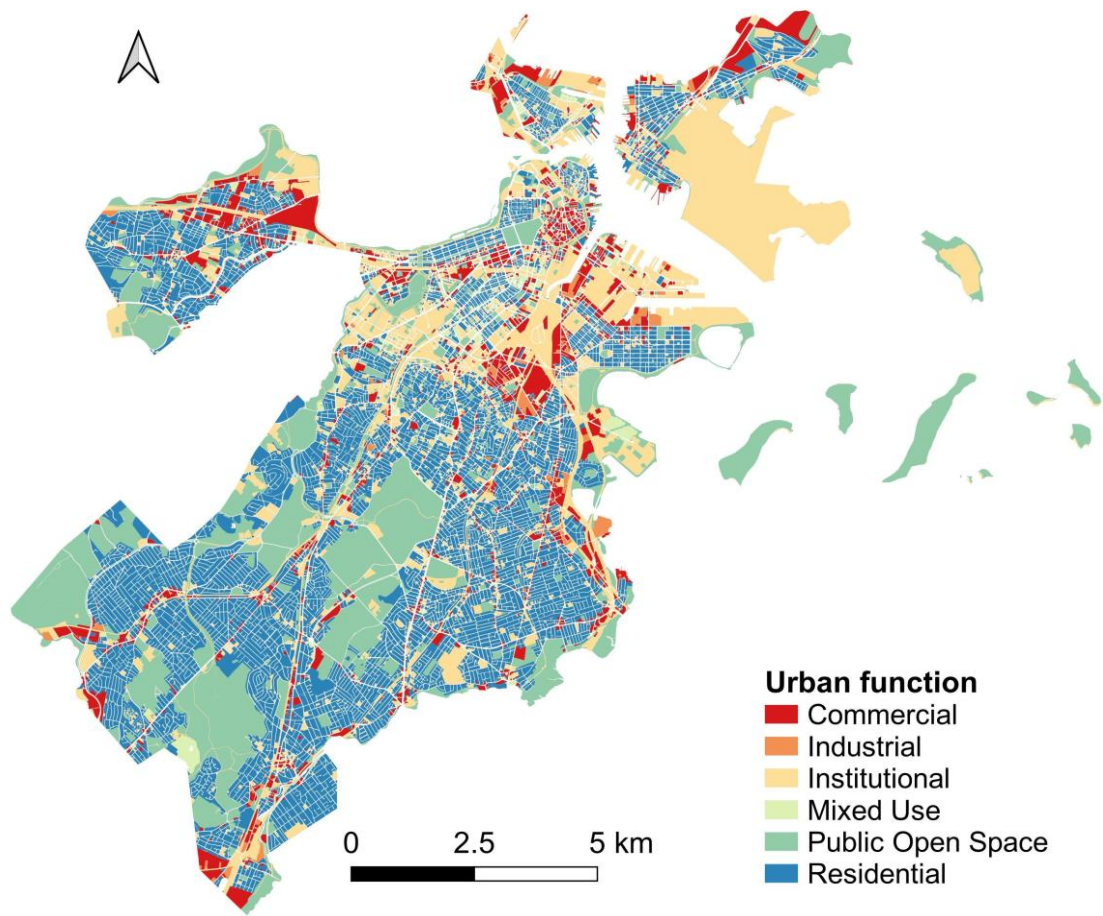

Figure 5 Boston land use at the block level

We used land use data of Boston at the block level and building footprint data in 2019. The block-level land use data in Boston is available at https://data.boston.gov/dataset/canopy-change-assessment-parcels-fy19-land-use. The land use data categorizes six types of urban functions in Boston, including commercial, residential, institutional, industrial, mixed used, and public open space functions (Figure 5). While the building footprint data are open access at OpenStreetMap (https://www.openstreetmap.org/). 98,331 buildings in Boston are collected from OpenStreetMap after data preprocessing (i.e., cleaning and fixing invalid geometries). 4,695 functional blocks are studied after filtering those with less than four buildings since the building configurations are the main target of this study of urban morphology. UTM zone 19N based on WGS 84 is employed as the projected coordinate system.

### 3.1.2 GNN model learned to predict urban function via morphology

As demonstrated in Table 2, the basic GNN model has been effectively trained, exhibiting consistent and satisfactory performance. Its weighted F1-score achieves 0.891 (±0.006). While its Micro F1-score achieves 0.907 (±0.005) with an overall accuracy of 0.908 (±0.009). They show the efficacy of the basic GNN model, as well as its capacity of generalization and stability. The Macro F1-score of 0.438 (±0.029) indicates the imbalanced performance in predicting different urban functions, as it does not take class imbalance into account and treats all classes equally. As shown in Table 3, the accuracies for different urban functions vary widely. A further discussion about the less satisfactory performance in several predictions is provided in detail in Section 4.1.2. Notwithstanding, the residential function is usually the main target in urban morphology research and thus this study focuses on the analysis of the residential function. The weighed metrics have higher priority than the other metrics in our consideration, as they reflect real-world class distribution. The weighted F1-score demonstrated the basic GNN model can stably predict urban functions via urban morphological graphs, particularly in predicting residential function (an accuracy of 0.988). In order to discover the correct knowledge learned from the GNN model, only the samples with correct prediction enter into the next step of interpretation.

Table 2 Performance of the GNN model in predicting urban functions. The average metrics with standard deviations are calculated across 10-fold results. While the aggregate metrics measure the combined results of the 10-fold test sets.

| Metric | Average (±SD) | Aggregate |
| --- | --- | --- |
| Weighted F1-score | 0.891±0.006 | 0.891 |
| Micro F1-score | 0.907±0.005 | 0.910 |
| Macro F1-score | 0.438±0.029 | 0.446 |
| Weighted Precision | 0.878±0.008 | 0.877 |
| Weighted Recall | 0.907±0.005 | 0.907 |
| Micro Precision | 0.907±0.005 | 0.910 |
| Micro Recall | 0.907±0.005 | 0.910 |
| Macro Precision | 0.448±0.043 | 0.452 |
| Macro Recall | 0.443±0.029 | 0.442 |

| | Overall accuracy | 0.908±0.009 | 0.907 |
|---|---|---|---|

Table 3 Confusion matrix of the aggregate performance generated by the original model not using random under-sampling.

| Pred\True | Commercial | Industrial | Institutional | Mixed use | Public open space | Residential |
|---|---|---|---|---|---|---|
| Commercial | 139 | 7 | 50 | 13 | 0 | 26 |
| Industrial | 0 | 0 | 0 | 0 | 0 | 0 |
| Institutional | 40 | 7 | 95 | 15 | 8 | 23 |
| Mixed use | 0 | 0 | 0 | 0 | 0 | 0 |
| Public open space | 3 | 4 | 10 | 0 | 36 | 1 |
| Residential | 72 | 6 | 68 | 75 | 8 | 3989 |
| Accuracy | 0.547 | 0.000 | 0.426 | 0.000 | 0.692 | 0.988 |

## 3.2 Analysis of core urban forms

### 3.2.1 Individual building-level core urban forms

This section aims to analyze the core urban forms in the individual-building level, i.e., the morphological patterns of core buildings. Figure 6 contains the key morphometrics (a), the symbolized morphological types of core buildings (b), and the comparison of the types (c). Figure 6 (a) indicates that shape is the main important dimension of building morphology for diverse urban functions. 10 important morphometrics are identified from the 22 used morphometrics, of which 8 are shape-related metrics covering the four types of urban functions. The core buildings in commercial functions are distinctive for their complexness (IPQ), compactness (Gibbs compactness), and elongation (elongation index) in building shape. Elongation is also characteristic of institutional buildings. The buildings in the public open space are characterized by their complexness (fractality), compactness (max circularity and DCM), and uniformity (BCI) of shape. While the morphological compactness and elongation are essential for residential buildings. The shape of the building footprint is significantly influenced by the way the land is used within the confines of the block. This has a direct bearing on the complexity of construction, the efficacy of utilization, and the user experience of the building.

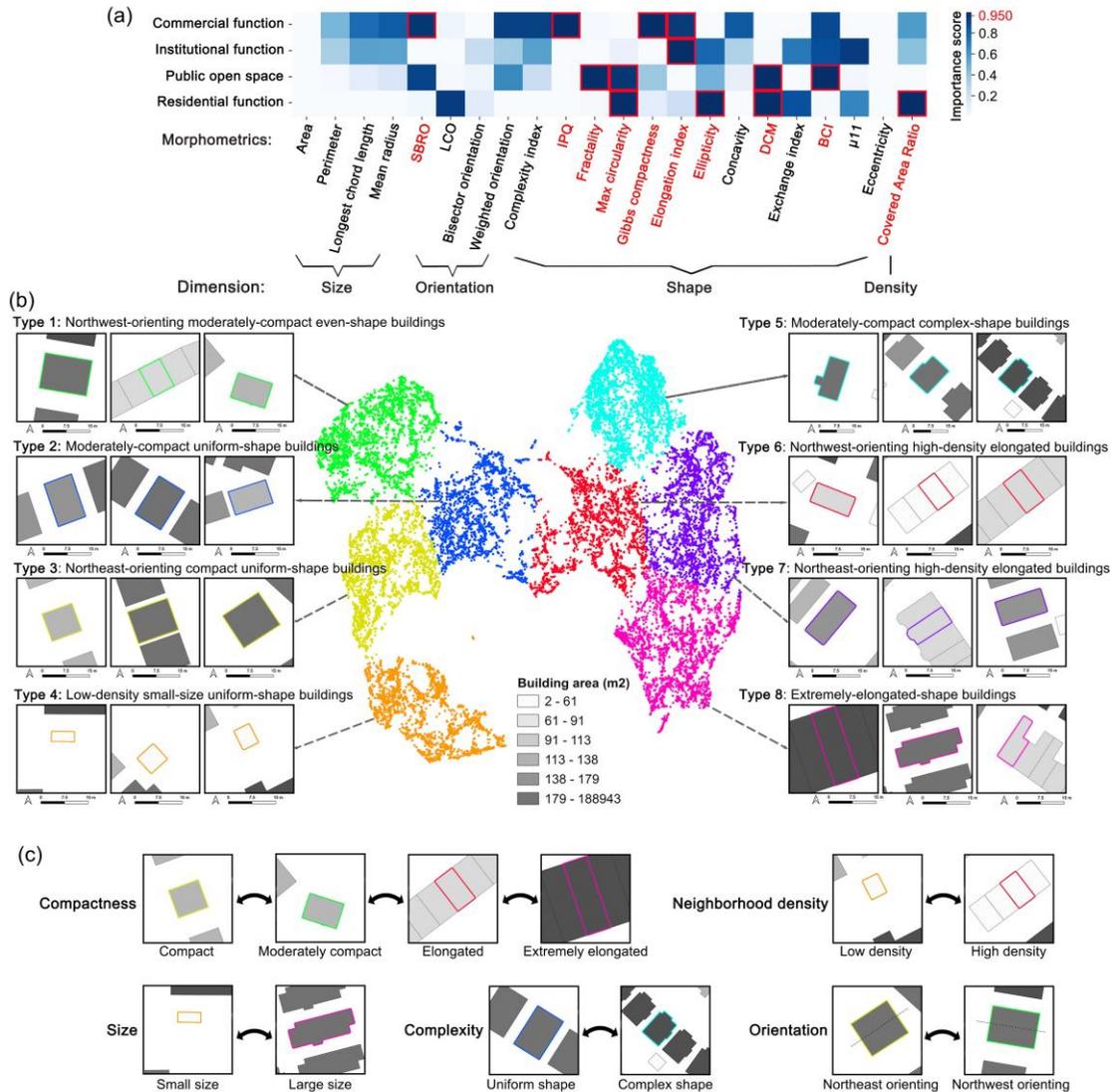

Figure 6 Analysis of core urban forms in the individual-building level. (a) The importance of morphological indicators for different urban functions. Those in red text are the most important metrics. (b) Eight types of core buildings shown in the low-dimensional representation space and their representative examples. (c) Examples of the different morphological descriptions shown in Panel (b).

Neighborhood density of buildings (covered area ratio) is a characterizing indicator for residential areas (Figure 6 (a)). It characterizes the distance of a building to the surrounding buildings. Thus, it is strongly related to the diverse living conditions, e.g., natural light exposure and window view satisfaction. A residential area with good livability usually has a very different neighborhood density from the other types of urban functions. On the other hand, the smallest bounding rectangle orientation (SBRO), an indicator of orientation, has a characterizing importance in commercial-use areas (Figure 6 (a)). It may be related to the visibility from primary thoroughfares and pedestrian pathways influencing customer flow. The urban spine of Boston runs from northeast to southwest, with the majority of its major arterials situated in this same direction. This may consequently influence the orientation of commercial buildings.

8 data-driven morphological types of core buildings are symbolized via the 10 key morphometrics (Figure 6 (b)). In the low-dimensional space, adjacent types are more similar in morphology. Thus, Type 4 is a distinctive type for its low density, small size and uniform shape. The remaining clusters are perceivable as two groups, one on the left and one on the right. The left one contains types with uniform shapes close to an ideal rectangle. Types 1-3 are different from each other for their compactness - moderately compact with a northwest-northeast orientation (Figure 6 (c)). While the right one consists of complex shapes that are considerably dissimilar to the ideal rectangle. A pattern was found that Types 5-8 are different in their compactness with an increasing elongation from top to bottom.

3.2.2 Block- & neighborhood-level core urban forms for residential function

Based on the 8 morphological types of core buildings, the analysis continues with an exploration of the block-level core forms (core building configurations) for residential function. In general, the statistics of the types of the residential core forms reveal a long-tail distribution (Figure 7 (a)). The head of the distribution is occupied by the dominant-type forms (42%) while the long tail consists of the diverse-type forms (58%). Amongst, the Type 5-dominant (T5-Dom) type of core forms is the most common. It consists of more than three Type-5 buildings and possibly one Type-4 buildings (Figure 7 (b)). Type 5 refers to the moderately compact and complex-shaped buildings, which is usually the form of single or two-family detached homes. Type 4 refers to the low-density small-sized uniform-shaped buildings, which is usually the form of auxiliary buildings for public services, e.g., waste collection points or storage points. The configurations combining Type 5 and Type 4 provide a historical and classical style of high-quality living environments, which was influenced by Garden city movement by Ebenezer Howard. Hence, this type of residential configurations is usually identified in the suburb areas of the Boston city (southwestern) for better environments and lower land prices (Cyan-blue areas in Figure 7 (c)).

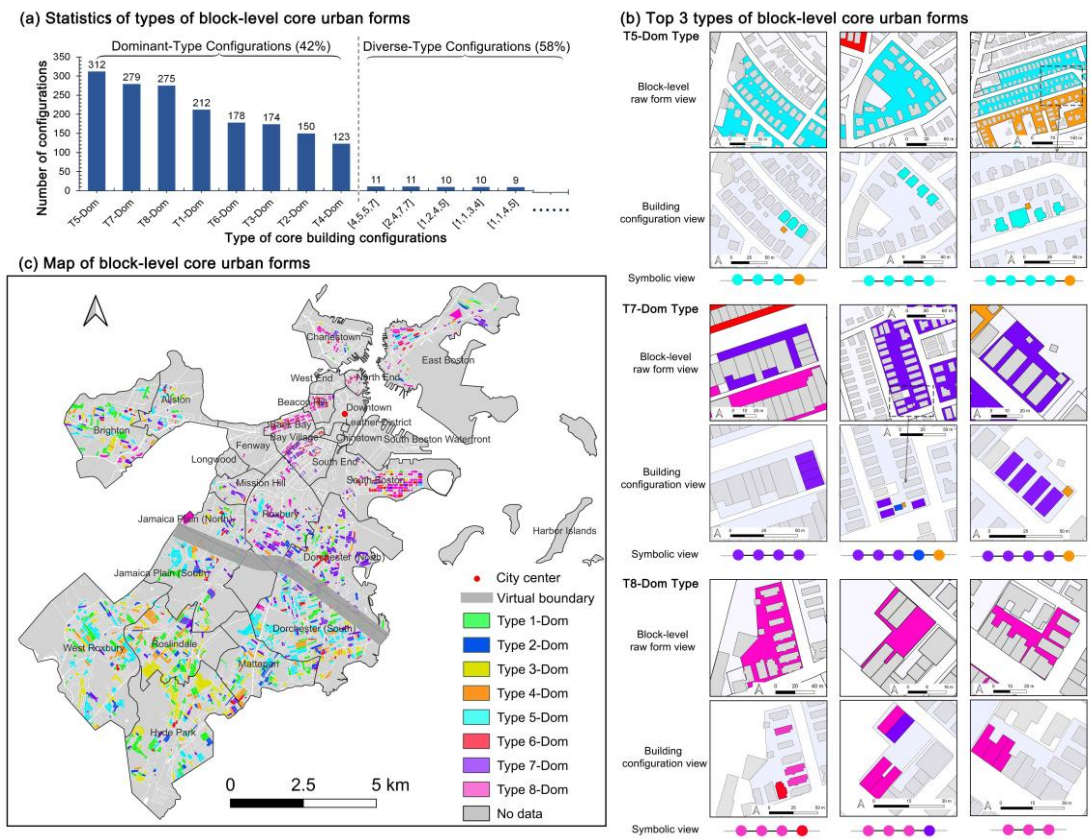

Figure 7 Block-level core urban forms for residential function. (a) Statistical numbers of different types of core forms. (b) Examples of the top 3 types of core urban forms, including a raw-form view, a core-building-configuration view and a symbolic view. (c) Geographic distribution of different types of block-level core forms. To better reveal the pattern, a virtual boundary (the grey transparent line) is drawn to separate Jamaica Plain, Dorchester, and Roxbury into the north and south parts.

Type 8-dominant (T8-Dom) and Type 7-dominant (T7-Dom) forms are ranked second and third in Figure 7 (a). They have been observed to frequently manifest in the city center and surrounding neighborhoods in Boston (Pink and purple areas in Figure 7 (c)). Specifically, the city center of Boston is located in the northeastern area, including Downtown, Beacon Hill, Back Bay, etc., which are surrounded by T8-Dom forms. The T8-Dom forms consist of more than three extremely elongated buildings (Figure 7 (b)). This extremely elongated shape can increase the usable area of the block land and accommodate more households due to the high land prices in the city center. Type 4 buildings, the auxiliary buildings, are not found in T8-Dom forms. Thus, necessary public services for the residents may be provided in a way of more efficient use of land in the areas. While T7-Dom forms are mainly located in the surrounding areas of the city center, which require a less high-density layout of the city center because of the slightly lower land prices. Thus, in T7-Dom forms, the dominant buildings (Type 7) exhibit a less elongated shape, while Type-4 buildings are utilized for public services. In summary, these representations of configurations reveal the high-order relationship between residential function and multiple urban forms.

In the neighborhood level, the geographical distribution of the regional representative core building configurations presents a gradual pattern along the urban spine of Boston (Figure 8 (a)). An imaginary line from northeast to southwest (the dotted line with arrows) is also drawn to portray the urban spine of the city to help analyze the regional representative core building configurations. It is evident that the representative types present a gradual geographic pattern along the urban spine. And we found that the pattern along the urban spine can also be discovered in the low-dimensional space from Type 8 to Type 3 (Figure 8 (c)), which represents that this pattern is also a gradual transition in urban morphology. Specifically, from northeast to southwest, the morphology of dominant core buildings transitions from the extremely elongated shape (Type 8) to the compact uniform shape (Type 3), illustrating the changes in the way different land use efficiencies are exploited (Figure 8 (d)). Regarding the urban spatial structure of Boston, the center-urban-suburban transition is also distributed from the northeast to the southwest along the urban spine. Hence, the pattern of neighborhood-level core forms along the urban spine fits the transition of center-urban-suburban.

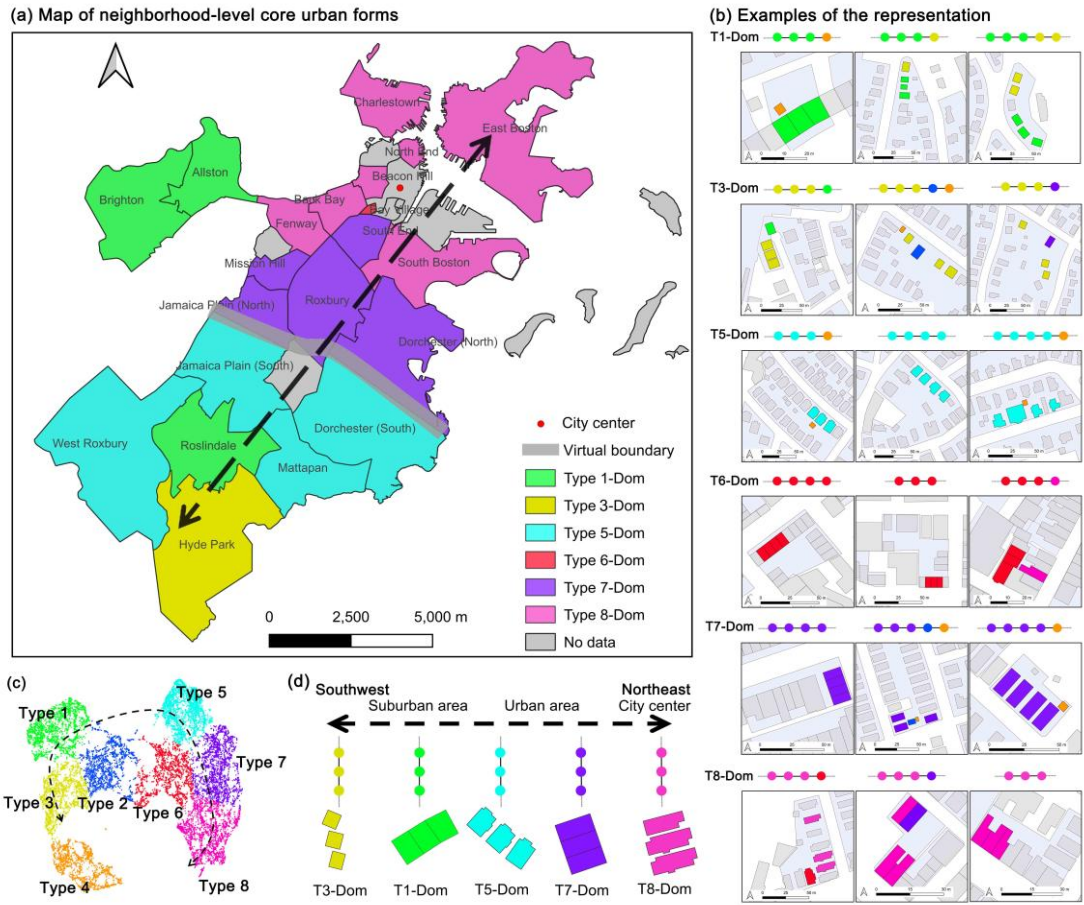

Figure 8 Neighborhood-level analysis of residential core forms. (a) Geographic distribution of regional representative types of residential core forms along the urban spine (the dotted line with arrows). (b) Three examples of the regional representative core forms in a core-building-configuration view. (c) Pattern along the urban spine projected in the low-dimensional space. (d) Examples of the core urban morphology representations for the regional representative types along the urban spine.

### 3.2.3 Quantitative analysis of residential core urban forms via land use efficiency

The aim of this section is two-fold. Firstly, to further demonstrate that the alteration of urban residential morphology fits the center-urban-suburban transition by influencing the land use efficiency, the relationship between land use efficiency and distance to the city center was analyzed. Two indicators of land use efficiency were calculated for residential blocks, i.e., the ratio of building area to block area ($E_{area}$) and the ratio of the number of buildings to block area ($E_{num}$). The regression analysis between distance to the city center and land use efficiency was fitted via the Power Function $y = ax^b$. The distance to downtown Boston is regarded as the distance to the city center. Secondly, to demonstrate the effectiveness of the core urban forms identified by the proposed CoMo framework, blocks with different types of urban forms were employed for calculating land use efficiency. Three groups of residential blocks were set up. (1) The group of residential blocks is represented with original complex urban forms. (2) The second group of residential blocks is represented with all core forms comprising any type of core building configurations identified by CoMo. (3) The third group of residential blocks is represented with the regional representative forms that analyzed in the neighborhood level (Figure 8).

The regression model in Figure 9 (e) proves the strong correlation between the distance to city center and land use efficiency. The $R^2$ value achieves 0.721 and all the parameters in the regression pass the significance test (Table 4). It indicates that, with a 99.9% confidence level, the distance to the city center can account for 72.1% of the observed variation in land use efficiency. Thus, the land use efficiency of residential blocks is inversely proportional to their distance from the city center. Concurrently, Figure 10 illustrates the evolving pattern of land use efficiency across the regional representative types along the urban spine. As the block is further away from the city center, the land use efficiency corresponding to the core building configuration tends to decline. It demonstrates that urban morphology directly affects land use efficiency. In summary, land use efficiency explains why the morphological pattern of core urban forms fits the center-urban-suburban transition.

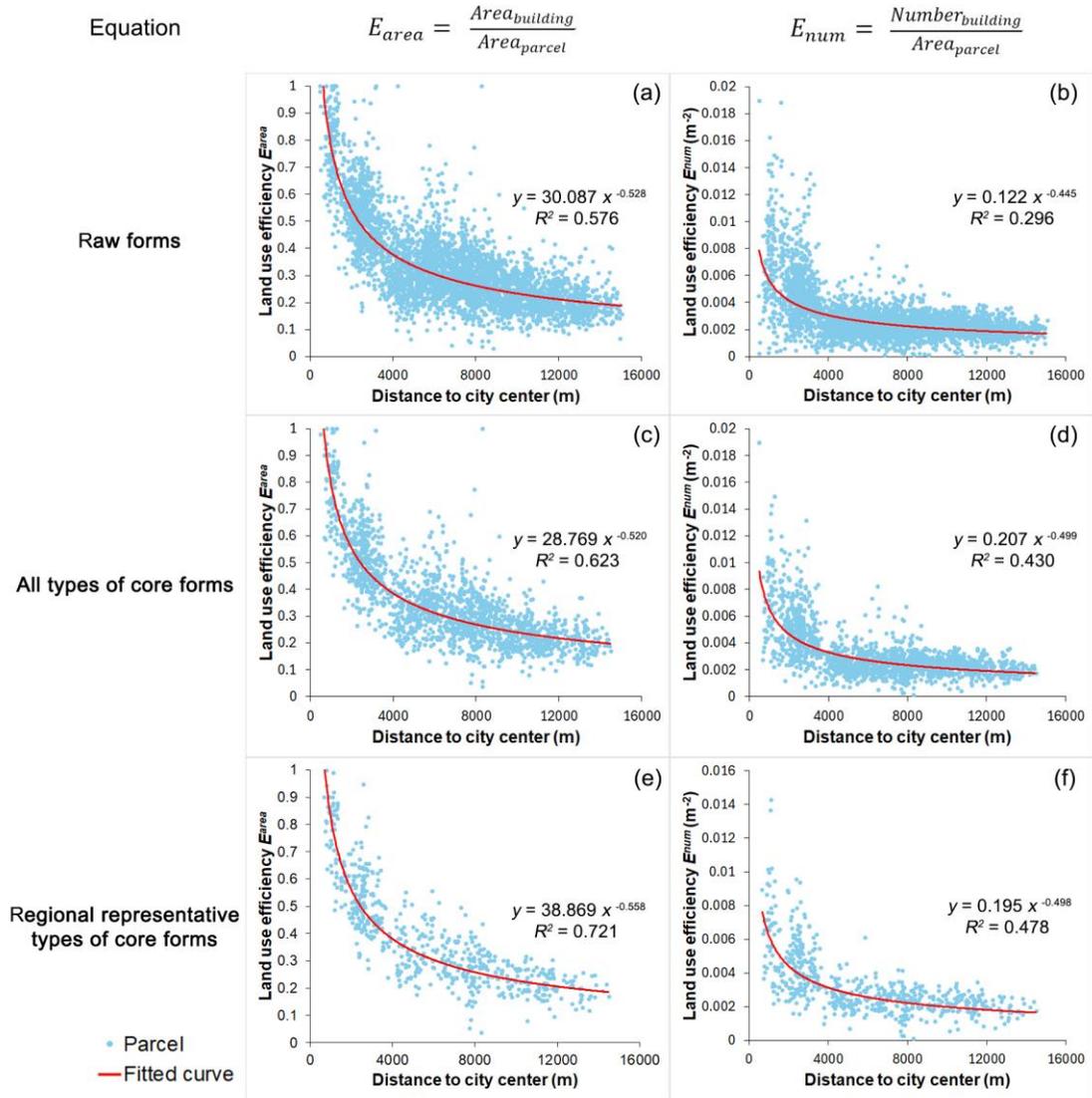

Figure 9 Comparison of the fitted relationships between distance to city center and land use efficiency calculated using different urban form representations and equations. (a)-(b) use the original form representation to calculate the two types of land use efficiency. (c)-(d) use all types of the core form representation to calculate the land use efficiency. (e)-(f) use the regional representative types of core form representation analyzed in Figure 8 for land use efficiency.

Table 4 Regression Results for the impact of distance to city center on land use efficiency

| log *y* | Raw forms | | All core forms | | Regional representative forms | |
|---|---|---|---|---|---|---|
| | log $E_{area}$ | log $E_{num}$ | log $E_{area}$ | log $E_{num}$ | log $E_{area}$ | log $E_{num}$ |
| $R^2$ | 0.576 | 0.296 | 0.623 | 0.430 | **0.721** | **0.478** |
| Adjusted $R^2$ | 0.576 | 0.295 | 0.623 | 0.430 | **0.721** | **0.477** |
| F-value | 5477.589 | 1693.756 | 2745.271 | 1254.368 | 1539.289 | 544.879 |
| p-value | <0.001 | <0.001 | <0.001 | <0.001 | <0.001 | <0.001 |
| Standard error | 0.126 | 0.191 | 0.119 | 0.169 | 0.111 | 0.167 |
| log $a$ | 1.478*** | -0.914*** | 1.459*** | -0.685*** | 1.590*** | -0.709*** |
| | (0.027) | (0.041) | (0.037) | (0.053) | (0.052) | (0.079) |
| $b$ | -0.528*** | 0.445*** | -0.520*** | -0.499*** | -0.558*** | -0.498*** |
| | (0.007) | (0.011) | (0.010) | (0.014) | (0.014) | (0.21) |

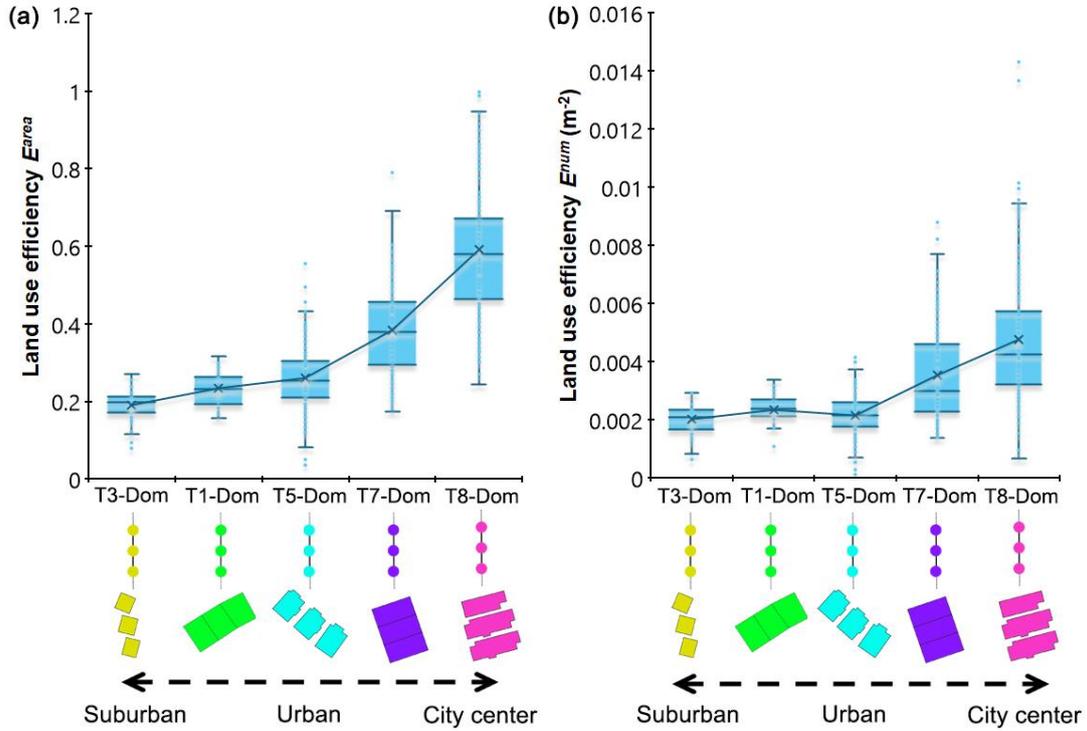

Figure 10 Land use efficiency of the regional representative types of core urban forms identified in Figure 8 (d). (a) Box plot of the indicator $E_{area}$ - the ratio of building area to block area. (b) Box plot of the indicator $E_{num}$ - the ratio of the number of buildings to block area.

The comparison of the regression results of the three groups in Figure 9 highlights the effectiveness of the core forms identified by CoMo. The $R^2$ values for the three rows in Figure 9 exhibit a notable increase ($E_{area}$: 0.576->0.721, $E_{num}$: 0.296->0.478). The regression model employing the regional representative forms (Group 3) demonstrated an improvement in the $R^2$ value of up to 0.182. And Table 4 demonstrates that all the six regression models are statistically significant. Hence, the proposed core urban morphology representation can effectively represent the complex urban forms, thereby helping explain the intrinsic relationship between urban form and function.

3.2.4 Core urban forms linked to other functions

Among the remaining three urban functions, it was observed that each function exhibited one or two dominant types of block-level core urban forms that occurred with high frequency (Figure 11 (a)(c)(e)). The commercial function is mainly reflected in T6-Dom and T8-Dom core forms, as shown in Figure 11 (b). Type-6 and Type-8 buildings are primarily defined by their elongated architectural forms, with no constraints on the uniformity of their shapes. Hence, complex shapes of buildings are frequently observed in commercial areas, which may be attributed to the optimization of land utilization. Type-6 buildings show a northwest orientation, which may be related to the necessity of the visibility from primary thoroughfares for commercial buildings. These thoroughfares frequently follow an orientation that is aligned with the northwest-orienting urban spine in Boston, a factor that is usually considered for commercial buildings.

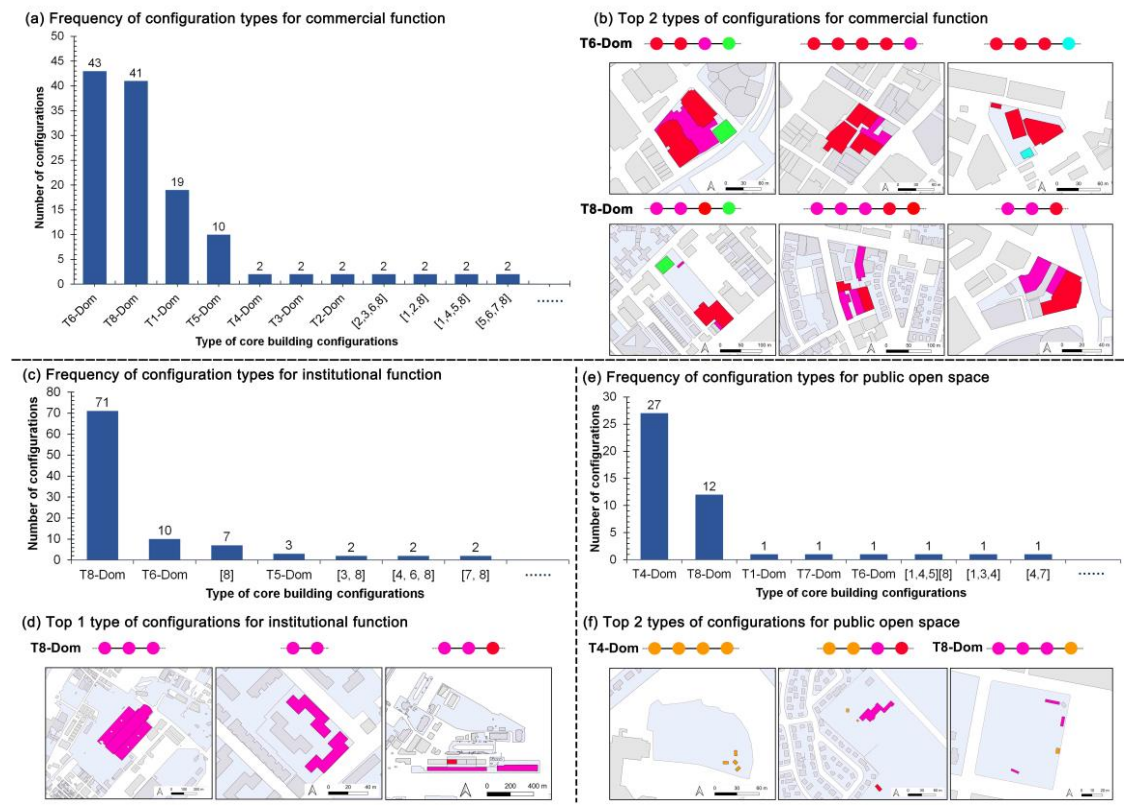

Figure 11 Types of core building configurations and examples for other functions, including commercial (a-b), institutional (c-d), and public open space (e-f).

T8-Dom forms occur often for the institutional function (Figure 11 (c)). The institutional buildings are for public service purposes, including exhibition centers, schools, terminals for transportation, etc. These structures are characterized by their elongated forms and intricate architectural details, often encompassing a multitude of office spaces (Figure 11 (d)). Thus, T8-Dom forms are the main urban form for institutional areas.

The core urban form of public open space mainly consists of Type-4 buildings, Type-8 buildings and their configurations (Figure 11 (e)(f)). Public open space can be regarded as a special type of institutional function. Thus, buildings in the public open space are also for public service purposes. They can be categorized as small buildings for the usage of public toilets, service stations, garbage collection points, etc. (Type-4 buildings), and large buildings for exhibitions and museums (Type-8 buildings). The T4-Dom form is a unique urban form for public open space among diverse urban functions, thereby making it more distinguishable than the commercial and institutional functions.

## 4. Discussion

### 4.1.1 AI-driven evidence for classic urban location theory

Our findings can contribute to the classic urban location theory by adding evidence to the relationship between location and urban morphology associated with urban functions. The location theory, first proposed in the early nineteenth century (Von Thünen, 1826) and later introduced for urban studies by Willian Alonso (Alonso, 1964), is the origin and foundation of digital twins models (Batty, 2024). Alonso explained the relationship between land prices and the location of urban spatial structure from an economic perspective. In the theoretical monocentric model (Figure 12 (a)), the land price increases significantly in close proximity to the city center, due to the higher level of investment potential in the center and the costs associated with commuting. Therefore, it influences citizens' demand and choice of location for residential purposes, thus forming the bid-price curve (Figure 12 (b)). According to our findings, we argue that the urban residential function alters the urban forms to enhance land use efficiency (Figure 12 (c)) in order to align it with the increasing land prices in locations closer to the center of the urban spatial structure (Figure 12 (b)). In the case of Boston, along the urban spine, the residential urban forms transform from the configurations of elongated buildings for high land use efficiency, to the configurations of moderately compact buildings for better living conditions (Figure 8 (c)(d), Figure 10). This morphological transition coins with the geographic center-urban-suburban transition of a city (Figure 8), significantly following a power function (Figure 9 (e), $R^2$=0.721, p<0.001). This pattern is the result of supply and demand balance, which proves Alonso's bid-price curve and the power function is also the form of urban scaling law (Bettencourt, 2013; Delloye et al., 2020). It can be used as evidence for the classic urban location theory. Furthermore, these AI-driven findings can support the construction of digital twins by effectively mastering the underlying mechanisms of complex city systems.

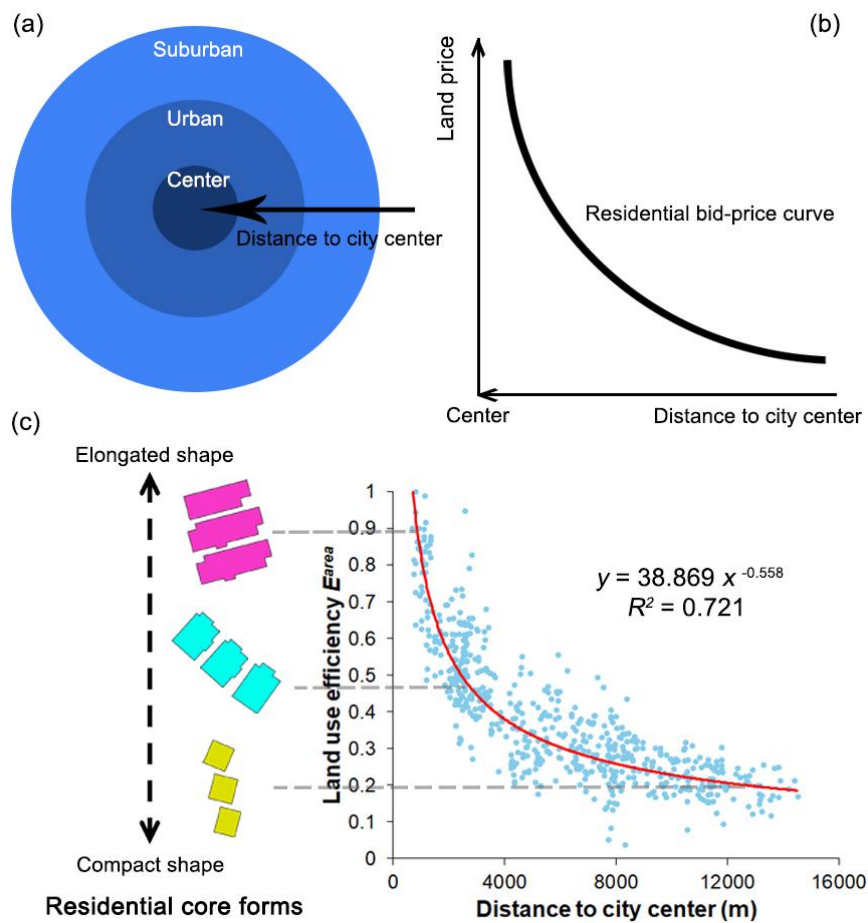

Figure 12 Classic urban location theories and the found evidence identified via CoMo. (a) The Alonso-Muth-Mills monocentric city model with the center-urban-suburban transition (Alonso, 1964; Mills, 1972; Muth, 1969). (b) Alonso's location-choice theory - residential bid-price curve (Alonso, 1964). (c) A significantly strong relationship between location and residential core forms represented by CoMo,

which matches Alonso's residential bid-price curve in (b).

### 4.1.2 Discussion on the less satisfactory performance of predicting several urban functions

The prediction of different types of urban functions shows different performances of the graph neural network model (Table 3). Specifically, three levels of performance are shown, ***good*** for residential and public open space functions (66-100), ***moderate*** for commercial and institutional functions (33-66), and ***poor*** for industrial and mixed-use functions (0-33). This phenomenon may be attributable to three potential factors, i.e., class imbalance, insufficient training samples and weak relationship between form and function. Firstly, in order to examine the impact of class imbalance, we followed a similar work (Izzo et al., 2022) and conducted an experiment to compare a basic GNN model that used a class-balancing strategy with the original model that used no class-balancing strategy. The details of the comparative experiment are provided in Section S3 (see *Supplementary Materials*). Table S2 shows that the original model outperforms the model that employs the class-balancing strategy on the majority of the evaluation dimensions. And the comparison of the confusion matrices of the two models indicates that the class balancing strategy can marginally enhance the prediction accuracy of some minority classes (Commercial: +0.004, Industrial: 0, Institutional: +0.004, Mixed use: 0->0.243, Public open space: +0.039, Residential: -0.065). However, these enhancements are insufficient to alter the levels of performance for any functions (Table S3 and Table S4). Hence, the class imbalance issue is not the main reason.

Secondly, the impact of insufficient training samples is reasoned based on the comparison of training sample sizes of different classes. Table S3 highlights the training sample sizes of different urban functions (see *Supplementary Materials*), which can reveal the minimum size of samples required for training the GNN model. Here, the sample size of public open space that produces a good prediction accuracy of 0.692 is taken as the minimum size of training samples. Only the sample size of the industrial function is smaller than the dividing threshold, which may be inadequate to meet the minimum requirement for GNN models. However, in contrast, the predictions for mixed-use, institutional, and commercial functions are reliable with sufficient training samples as their sample sizes exceed the minimum size. In summary, it is hypothesized that the accuracies of the three functions are a result of their relationships with urban morphology. We argue that urban morphology exhibits a weak relationship to the mixed use function, while a moderate correlation with institutional and commercial functions.

### 4.1.3 Limitation and outlook

The CoMo framework is predicated on the assumption that the input building footprint data have perfect quality. In this study, a manual check was conducted first to assess the OSM data quality in Boston, thereby confirming the data homogeneity across the city. Our supplementary experiment showed that a 2-meter error results in a minor decline of approximately 0.005 in the F1-score for GNN, whereas a 5-metre error causes an acceptable decrease of approximately 0.015 (See Section S4 in *Supplementary materials*). Notwithstanding, it is recommended that relevant urban morphology researchers undertake quality checks on building data.

Two limitations will be further explored. Firstly, our CoMo framework does not yet account for the potential influence of external factors as it is currently focused on the study of the relationship between urban morphology and urban function, the two fundamental aspects of urban systems. Thus, the input of the basic deep learning model is all morphology-related data. In the reality, however, the intricate and multifaceted nature of urban systems encompasses a multitude of factors, some of which have the potential to influence the interrelationship between urban form and function. In the future, CoMo will incorporate multi-source data to comprehensively portray the intricate and multifaceted urban systems and find out potential factors. Secondly, CoMo was only implemented in Boston, a classic city in urban morphology studies. Our future work will evaluate the performance of CoMo in more cites and reveal the urban mechanisms at a large geographic scale.

## 5. Conclusion

Urban researchers have been endeavored to construct mathematical and computable models of complex urban cities with the objective of facilitating sustainable urban development (Batty, 1971). The concept of digital twins has been proposed as a paradigm of quantitative models of simulating urban dynamics (Batty, 2024), so that urban planners can intervene in advance in the urban form based on digital twins to circumvent issues that impede urban resilience and sustainability (Batty, 2024). Furthermore, to better model the mechanisms of urban systems, it is important to reveal the high-order relationship between urban form and function in an accurate and explicable way. Nevertheless, existing methods have difficulty in accurately modeling the complex urban forms while simultaneously interpretating them explicably in human terms.

This paper proposed the CoMo framework to symbolize the complex urban forms into the explicable representation of core urban morphology for better interpreting the links to urban functions. The proposed method follows a three-step framework of "graph construction - learning - interpretation" to reveal the relationship between urban form and urban function from a deep learning GNN model. It differs from other methods in two main ways: (1) CoMo represents complex urban forms with the core urban morphology representation, providing a new perspective for urban morphology-related studies. In particular, the core urban morphology representation has two advantages: (i) the spatial-graph pipeline characterizes the complex spatial connections between buildings and their spatial configurations. (ii) It interprets the complex urban forms as readily explicable spatially symbolic patterns for better human understanding, which are essential for city modeling (Batty, 2024) as they simultaneously convey complex mechanisms. (2) The high-order relationship between urban form and urban function is revealed by interpreting the explainable graph neural network. CoMo symbolizes continuous variables of morphological metrics and matches them with discrete variables of urban functions, thus facilitating a novel way of relationship analysis and comprehension. Furthermore, the specific module in the proposed framework is extensible. The GCN model can be replaced by other GNN models, such as Graph Attention Network (Veličković et al., 2018) and GraphSAGE (Hamilton et al., 2017). The technologies in the dimensional reduction part and clustering part are also replaceable.

An experiment in Boston, USA, was conducted as a case study. The stable weighted F1-score of 89.14% for predicting urban function indicates that urban morphology is strongly related to urban function and CoMo accurately learned the relationship between form and function. Some findings of urban morphology are revealed for the selected study area - the city of Boston. (1) 8 data-driven types of individual building-level core forms are identified. Shape is a characterizing dimension of building morphology for diverse urban functions because the shape of elongation-compactness and uniformity-complexity can directly affect land utilization efficiency, construction cost and building user experience. Neighborhood density is important for residential buildings, whereas orientation characterizes commercial buildings. (2) The block-level residential core forms indicate that the forms dominated by elongated buildings tend to be located near the urban center to increase land use efficiency due to the high land prices near the center. By contrast, the urban forms far from the urban center are mainly composed of the configurations of moderately compact buildings and a few auxiliary buildings to provide better living conditions. (3) Neighborhood-level residential core forms are affected by urban spatial structure because the morphology follows a geographic center-urban-suburban transition. A significantly strong relationship is revealed between location and residential core forms, which matches the residential bid-price curve, thus proving Alonso's classic urban location theory. (4) The elongated complex buildings dominate commercial and institutional forms. While small, low-density compact buildings are more common in public open spaces.

In summary, different urban functions are associated with different urban forms, revealing the high-order relationship between function and morphology. These findings help enrich the relevant theories and evidence for city modeling and digital twins, thus providing reference for urban planners to intervene in urban systems and promote sustainable urban development.

# Acknowledgement

This work was supported by the German Research Foundation (DFG) [500249124]